\documentclass[sageh,times]{sagej}
\usepackage{moreverb,url}
\usepackage[colorlinks,bookmarksopen,bookmarksnumbered,citecolor=blue,urlcolor=blue]{hyperref}
\usepackage{verbatim}
\newcommand{\note}{\textcolor{red}}  
\newcommand\BibTeX{{\rmfamily B\kern-.05em \textsc{i\kern-.025em b}\kern-.08em
T\kern-.1667em\lower.7ex\hbox{E}\kern-.125emX}}
\def\volumeyear{2020}
\begin{document}

\runninghead{False positives in social cognitive mapping}

\title{False positives using social cognitive mapping to identify children's peer groups}

\author{Zachary P. Neal\footnote{Psychology Department, Michigan State University}, Jennifer Watling Neal$^1$, and Rachel Domagalski\footnote{Mathematics Department, Michigan State University}}

\corrauth{Zachary Neal, Psychology Department, Michigan State University, East Lansing, MI 48824, USA; zpneal@msu.edu}

\begin{abstract}
\textbf{\note{This pre-print has been relocated to the PsyArXiv server. Subsequent revisions can be found at \href{https://psyarxiv.com/yfmzd/}{https://psyarxiv.com/yfmzd/}}.}\\
\newline
Children and adolescents interact in peer groups, which are known to influence a range of psychological and behavioral outcomes. In developmental psychology and related disciplines, social cognitive mapping (SCM), as implemented with the SCM 4.0 software, is the most commonly used method for identifying peer groups from peer report data. However, in a series of four studies, we demonstrate that SCM has an unacceptably high risk of false positives. Specifically, we show that SCM will identify peer groups even when applied to random data. We introduce backbone extraction and community detection as one promising alternative to SCM, and offer several recommendations for researchers seeking to identify peer groups from peer report data.
\end{abstract}

\keywords{Accuracy, Peer groups, Social networks, Social cognitive mapping, Validity}

\maketitle

\section{Introduction}
Decades of research demonstrate the importance of peers for child and adolescent development and psychological well-being~\citep{bukowski2018,gifford2003}. Children and adolescents interact in peer groups with structural and behavioral features that are associated with a wide range of psychological, social, and academic outcomes~\citep{birkett2015,espelage2003,ryan2001}. However, identifying peer groups can be challenging and represents a critical measurement task for developmental and clinical researchers~\citep{kindermann2018}. To overcome these challenges, Cairns and colleagues proposed social cognitive mapping (SCM), a method of peer group identification that involves identifying peer groups using multiple peer reports of groups of children that interact together in a setting such as a classroom~\citep{cairns1994,cairns1988}.

SCM has become a dominant method for identifying children's peer groups from peer report data. The data are easy to collect, the ability to triangulate from peers reduces the impact of non-response, the analysis is easy to perform, and there is some evidence for its validity~\citep{gest2003}. However, nothing is known about the extent to which SCM can yield false positives, where the method identifies peer groups from data that contain no or only weak evidence of their existence. In this paper, we show that SCM has an high rate of false positives, assigning on average two-thirds of children to peer groups even when it is applied to random peer report data. We conclude that researchers should \emph{not use SCM, particularly as it is implemented in the \texttt{SCM 4.0} program, to identify peer groups}, and should explore alternative methods for identifying peer networks and peer groups from peer report data~\citep{neal2014}.

We begin by reviewing SCM, providing an overview of its origins, where and how it has been used, how it works, and evidence for its accuracy. Then, in a series of four related studies, we confirm that SCM can detect true positives, examine SCM's risk of false positives under different conditions, and explore backbone extraction and community detection as an alternative to SCM. In the discussion section, we synthesize the key findings from these studies, offering recommendations for researchers seeking to identify children's' peer groups.

\section{Background}

\subsection{What is social cognitive mapping?}
Experiences in peer groups play a significant role in childhood and adolescent development~\citep{howe2010,kindermann2018,rubin2015}. Specifically, aspects of peer group structure (e.g., size, hierarchy) or behavior (e.g., norms) have been linked to psychological (e.g., depression), social (e.g., aggression, homophobic name calling, prosocial behavior, resource control) and academic (e.g., motivation, achievement) outcomes~\citep{birkett2015,espelage2003,ryan2001,zhao2016,zarbatany2019}. However, identifying peer groups presents a range of challenges in developmental studies. For example, self-report methods are sensitive to self enhancement bias and missing data, while observational methods are resource intensive~\citep{cairns1994,kindermann2018,neal2013a}. To overcome some of these challenges, Cairns and colleagues proposed SCM as an alternate method for the identifying of peer groups~\citep{cairns1988,cairns1994}. SCM relies on peer informants to provide reports of groups of children in a particular setting, such as a classroom, that hang out together. Through a series of aggregating and filtering transformations, SCM uses these peer-reported data to identify peer groups~\citep{neal2013a}. Specifically, SCM is intended to answer two questions: \emph{first}, do the children in this setting interact with one another in peer groups, and \emph{second}, if so, which children are members of which groups?

SCM developed in roughly three phases. First, during the \emph{development} phase in the late 1980s, a research team at the University of North Carolina at Chapel Hill led by Robert Cairns and Beverley Cairns experimented with ways to triangulate multiple children's reports of peer groups into a single picture of a setting's social structure and its peer groups~\citep{cairns1985,cairns1988,cairns1989}. This work concentrated on examining a matrix of children's co-occurrence in reported peer groups, evolving from ``a decision rule procedure [in which] arbitrary standards were adopted'' to a more objective set of steps ``with minimal reliance on intuitive judgements''~\citep[][p. 817]{cairns1988}. Second, during the \emph{formalization} stage in the early 1990s, these steps were refined into a consistent procedure that appeared in an essentially identical form across multiple papers that included fully-worked examples~\citep{farmer1991,farmer1993,cairns1994}. This phase also included the development of software to facilitate the use of SCM~\citep{leung1998}.\footnote{Usually \textsf{SCM 4.0} is attributed to Leung only, however here we cite both Leung who wrote the manual and original program, and Alston who is identified in the software itself as the programmer.} Finally, the \emph{application} phase from the mid-1990s onward has involved the use of SCM throughout developmental psychology and related fields focused on studying children's peer relations and groups, as well as the development of variations on the steps developed during the formalization stage. In particular, two of Cairns' colleagues developed their own variations: one relying on conditional probabilities and a binomial $z$ test~\citep{kindermann1993}, and another relying on principal components analysis~\citep{gest2007}.

\subsection{How often is SCM used?}
To determine the specific variant of SCM that is most commonly used, we started with a dataset of 201 papers from a recent review of social network data collection methods in developmental psychology~\citep{neal_data}. These papers were initially identified using Google Scholar and reflect papers published or online (a) prior to February 2019 (b) in the 30 top-ranked journals classified by Web of Science as ``Psychology, developmental'' in 2016 (c) that contained the phrase ``social network'' and one or more network-relevant keywords (e.g. density, centrality, clique, etc.)~\cite{neal_inpress}. We reviewed each paper and identified 73 that attempt to identify network-based peer groups or cliques. A majority of these papers (N = 46, 63\%) used SCM to identify peer groups. Among those using SCM, most (N = 38, 83\%) used the specific variant described by~\cite{cairns1994}.\footnote{Of the remaining papers, 3 use the variant described by~\cite{kindermann1993}, 2 use the variant described by~\cite{gest2007}, and 3 provided insufficient detail to determine the variant.} Finally, among the papers using Cairns and Cairns' version of SCM, nearly half (N = 16, 42\%) explicitly noted that they used \textsf{SCM 4.0}, a DOS-based program that implements this version of SCM. Because it is the most widely-used method, in this paper we focus on SCM as described by~\cite{cairns1994} and as implemented in \textsf{SCM 4.0}~\citep{leung1998}.

To determine how widely SCM is used and the extent to which it is used outside developmental psychology, where it wad first developed, we examined each paper citing~\cite{leung1998}. Using Google Scholar we located an additional 26 papers appearing in such youth-focused fields as school psychology~\citep{farmer2010}, social psychology~\citep{wolfer2012}, special education~\citep{avramidis2010}, STEM education~\citep{radovic2017}, and substance use~\citep{sheppard2012}. We also observed that it is used outside North America, by researchers in Latvia~\citep{levina2012}, Korea~\citep{ahn2011}, Norway~\citep{fandrem2010}, and Spain~\citep{bacete2013}.

Combining the results from~\cite{neal_inpress} and our own search, we identified a total of 42 papers using SCM. They were published in such flagship journals as \emph{Developmental Psychology} and \emph{Child Development} between 1995 and 2019 (M = 2009.6, SD = 5). We therefore conclude that SCM is among the most widely- and currently-used methods for identifying peer groups.

\subsection{What is SCM used to study?}
Researchers use the peer groups identified by SCM in multiple ways. First, some researchers use SCM-derived peer groups to generate group-level behavioral norms, then estimate mixed models to examine associations between these group norms and individual psychological and social outcomes~\citep{chung2010, zhao2016}. For example,~\cite{zhao2016} found that children who participated in SCM-derived peer groups with higher levels of average social withdrawal exhibited less social competence, less positive school attitudes, and higher levels of depression. Second, some researchers study the association between compositional (e.g., ethnic composition) or organizational features (e.g., hierarchization) of SCM-derived peer groups and psychological or social outcomes~\citep{shi2014, zarbatany2019}. For example, \cite{shi2014} found that the socialization of aggression differed depending on the ethnic composition of SCM-derived peer groups. Finally, some researchers have used SCM-derived peer groups to examine the extent to which teachers are accurate observers of classroom peer relationships (i.e. teacher attunement)~\citep{gest2006, hoffman2015}. For example, \cite{hoffman2015} found that elementary school teachers' reports of classroom peer groups exhibit only modest attunement to SCM-derived peer groups. 

\subsection{How does SCM work?}
SCM begins by collecting peer reports by asking participating children a question like \emph{Are there people in school who hang around together a lot? Who are they?} Each participating child is permitted to report any number of ``hanging around'' groups, and each group they report can contain any number of children including themselves. For example, Child A might report the existence of a ``hanging around'' group composed of children A, B, and C (report 1), and another group composed of children W, X, Y, and Z (report 2). Additionally, Child B might report the existence of a group composed of children A, B, C, and D (report 3). Thus, different reporters may report different numbers of groups (e.g. Child A provided two reports, while Child B provided one report), and these reports may partially overlap (e.g. Child A's first report and Child B's first report overlap). Then a series of aggregations and transformations are applied to these raw data to define a peer network and identify peer groups.

First, these peer report data are organized as a setting-wide ``recall matrix'' $\mathbf{R}$ that contains a row $i$ for each child in the setting and a column for each report $j$, so that each cell in the matrix $R_{ij}$ contains a 1 if child $i$ appeared in report $j$, and otherwise is 0.\footnote{The recall matrix does not contain information about \emph{who} provided any given report; this information is not used by SCM when identifying peer groups.} Returning to the example above, cell $R_{B1} = 1$ and cell $R_{B3} = 1$ because child B appeared in both reports 1 and 3, but $R_{X1} = 0$ because child X did not appear in report 1. The \textsf{SCM 4.0} software imposes some restrictions on the recall matrix that are not necessarily required by SCM in general: it can only contain data on up to 2000 peer-reported groups and up to 400 distinct children, and each peer-reported group can contain up to 20 members.

Second, the reports in the recall matrix $\mathbf{R}$ are transformed into a symmetric ``co-occurrence'' matrix $\mathbf{C}$ using
\begin{equation}
    \mathbf{C} = \mathbf{RR}'
\end{equation}
where $C_{ij}$ and $C_{ji}$ contain the number of times child $i$ and child $j$ were reported to be in the same group, and $C_{ii}$ contains the number of times child $i$ was reported to be in any group. In this step, SCM mirrors a classic example from the social network literature in which children's potential interactions are inferred from their co-participation in school clubs using bipartite projection~\citep{breiger1974}.

Third, $\mathbf{C}$ is transformed into a ``similarity'' matrix $\mathbf{S}$ using
\begin{equation}
    \mathbf{S} = cor(\mathbf{C})
\end{equation}
where $S_{ij}$ is the Pearson correlation coefficient of child $i$'s and child $j$'s column (or row) in $\mathbf{C}$. In this step, SCM mirrors the CONvergence of iterated CORrelations (CONCOR) algorithm for group detection~\citep{breiger1975}. However, unlike CONCOR, which repeatedly computes the correlation of the matrix (e.g. cor(cor(cor(\textbf{C})))), SCM performs this operation only once.

Fourth, a binary peer network $\mathbf{N}$ is constructed by defining child $i$ and child $j$ as connected if $S_{ij} \geq T$, where $T$ is a cut-off threshold between 0 and 1. \cite{cairns1994} recommended using a threshold value of 0.4, and this is the most commonly reported value used in the 42 papers we located.\footnote{Studies that included Wendy Ellis as an author always used 0.5; we do not use it here because it is generally not used by other authors and because it fails to correctly identify peer groups even in~\cite{cairns1994} original data (see study 1 below). Studies that included Thomas Farmer as an author sometimes, although not always, report using a `significant' threshold; we cannot use it here because these studies do not report what `significant' means in this context, how the threshold was identified, or the actual value of the threshold.}

Finally, groups of peers are identified from the peer network described by $\mathbf{N}$. These groups are permitted to overlap so that a single child may be a member of none, one, or more than one peer group. \cite{cairns1994} do not describe a specific method, and seem to suggest that the identification of groups in $\mathbf{N}$ is a trivial task that can be performed by visual inspection. At least two different descriptions of the method used to identify peer groups appear in the literature. \cite{farmer1993} and others report that it identifies peer groups so that each member of the group is connected to ``at least 50\% of the members in the cluster'' ~\citep[p. 234; see][]{avramidis2010,fandrem2010,rodkin2006}. Separately, \cite{bacete2013} report that it identifies peer groups by adding members to a group until no one ``is found who has a correlation profile equal to or greater than r = .40 with any of the members who have previously been incorporated into the group'' ([translated] p. 64-65). However, by examining the output generated by \textsf{SCM 4.0} in the analyses described below, we have verified that it does not use either of these methods for identifying peer groups from a peer network. Thus, we (and, seemingly, others) do not know exactly how SCM identifies peer groups from a peer network.

All five of the steps involved in SCM are automated by the \textsf{SCM 4.0} program~\citep{leung1998}, which has been used in at least 42 published studies to identify peer groups. Although we do not know exactly how \textsf{SCM 4.0} performs the final `group identification' step, in our analyses below we simply use the output generated by the program.

\subsection{Is SCM accurate?}
Because all methods measure social phenomena with error, the goal of measure development is to understand and minimize those errors. Therefore, it is often important to ask whether a given measurement approach is ``accurate.'' When SCM is used to study peer groups, there are at least three distinct varieties of accuracy that might be investigated.

\begin{center}
\begin{figure}
\includegraphics[width=.5\textwidth]{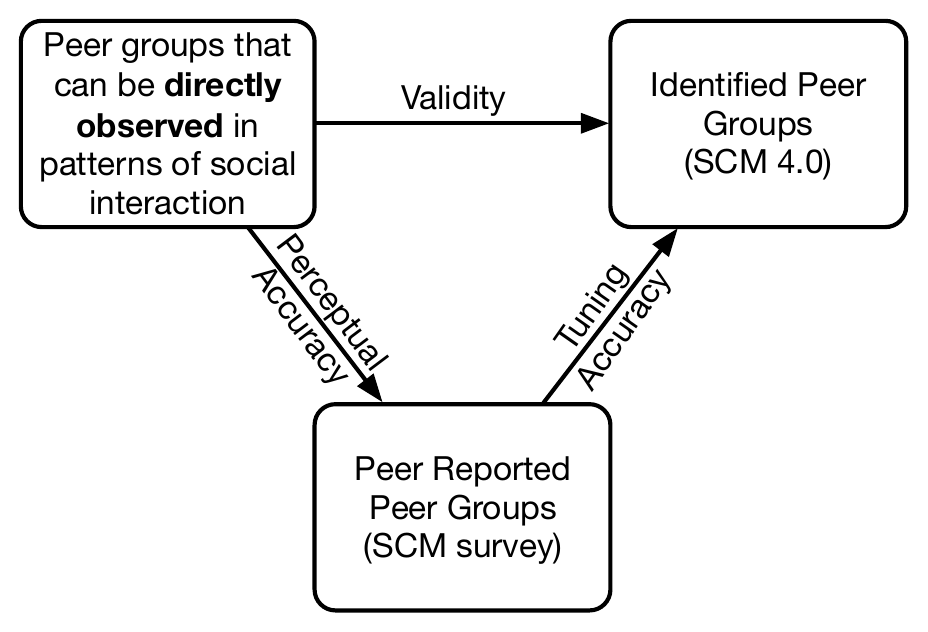}
\caption{Types of accuracy relevant when social cognitive mapping is used to identify peer groups from peer report data.}
\label{error}
\centering
\end{figure}
\end{center}

First, \emph{validity} describes the extent to which peer groups identified via SCM match peer groups that can be directly observed in patterns of children's social interactions. Because such observational data are difficult to collect~\citep{gest2003}, there have been limited attempts to evaluate SCM's validity. One early study of a 7\textsuperscript{th} grade classroom of 26 children, observed that the children were ``more likely to interact with members of their own [SCM-identified] subgroup than with members of other [SCM-identified] clusters''~\citep[][p. 107]{cairns1994}. Similarly, a larger study of 72 children in 4\textsuperscript{th} through 7\textsuperscript{th} grade found that ``Children were observed to interact with members of their SCM‐identified social cluster at a rate four times higher than with other same‐sex classmates''~\citep[][p. 513]{gest2003}.

Second, \emph{perceptual accuracy} describes the extent to which the peer groups identified by a particular child on an SCM survey and that appear in the recall matrix match the peer groups that can be directly observed. Because ``respondents have only a limited knowledge of the classroom'' it is assumed that ``there are frequent errors'' in the reports that each child provides, and thus that perceptual inaccuracy is common~\citep[][p. 104]{cairns1994}. For example,~\cite{neal2016} found that children have inaccurate perceptions about which of their classmates hang out together ($\kappa = 0.371$), but that girls, older children, and children in smaller classrooms were more accurate. Such inaccurate perceptions can still be informative, for example as an indicator of a child's social awareness~\citep{cappella2012}, and because an individual's behaviors are shaped by their perceptions of reality as much as by reality itself \citep{krackhardt1987}. However, SCM aims to overcome the expected perceptual inaccuracies of individual children by pooling and triangulating multiple children's reports.

Finally, a third type of accuracy which we call \emph{tuning accuracy} describes the extent to which the peer groups identified by SCM accurately summarize or triangulate the information contained in the peer reports collected via an SCM survey. To use the analogy of a radio tuner, the peer report data is a combination of signal (i.e. information about directly observable peer groups) and noise (i.e. random error due to the children's perceptual inaccuracies)~\citep{shannon1963}. The purpose of SCM, like a tuner, is to filter out the noise to yield a clear signal. There are two ways that SCM might exhibit tuning accuracy: first it can identify peer groups for which the peer reports contain evidence (true positives), and second it can fail to identify peer groups for which the peer reports do not contain evidence (true negatives). There are also two ways that SCM might exhibit tuning inaccuracy: first it can identify peer groups for which the peer reports do not contain evidence (false positives; type I error), and second it can fail to identify peer groups for which the peer reports do contain evidence (false negatives; type II error).

Each of these forms of accuracy is important. However, perceptual accuracy and validity are both challenging to establish because collecting observational data from many diverse settings would be cost- and time-prohibitive. In contrast, tuning accuracy can be evaluated without observational data by using simulated data as we describe below. Moreover, tuning accuracy is a critical prerequisite for validity. Without the ability to tune in the signal, and tune out the noise, no amount of perceptual accuracy will allow SCM-identified peer groups to be valid. Therefore, in this paper, we are interested in investigating SCM's tuning accuracy, and specifically its risk of false positives.

Three prior studies offer some insight into SCM's risk of false positives. First,~\cite{watts2003} explains that when a recall matrix is transformed into a co-occurrence matrix using Equation 1 ``even a random [recall matrix] -- one that has no particular structure built into it at all -- will be highly clustered'' (p. 128). Second,~\cite{pijl2011} found that compared to identifying peer groups from a network of reciprocated self-reported friendships, SCM assigned all children to a peer group, which was ``quite surprising, as it is known from the literature that 4-10\% of children do not have friends in primary classrooms'' (p. 484).  Finally,~\cite{neal2013a} demonstrated using illustrative data that ``distinct peer groups always appear to be present, no matter what responses children give'' during data collection (p. 605). Guided by these past studies, \emph{we hypothesize that SCM has a high rate of false positives}, identifying peer groups even from peer report data that lack evidence of peer groups. To investigate this hypothesis, we report on four separate studies: The first study confirms that SCM can detect true positives, the second and third studies examine SCM's risk of false positives under different conditions, and the fourth study explores backbone extraction and community detection as an alternative to SCM for identifying peer groups.

\section{Study 1: True positives in a benchmark classroom}
\subsection{Methods}
To examine whether SCM is able to detect true positives, correctly identifying when children are members of peer groups, we use data described by~\cite{cairns1994}. These data consist of a set of 61 peer reports collected from 17 children (11 girls, 6 boys) in a 26 child (15 girls, 11 boys) 7\textsuperscript{th} grade classroom.\footnote{The classroom contained 27 children, but one child (Pam) never appeared in any of the peer reports, and therefore is excluded from these analyses.} They offer an ideal benchmark dataset for two reasons. First, these are the data originally used by~\cite{cairns1994} to demonstrate and validate SCM, and therefore ought to offer SCM the best opportunity for correctly identifying peer groups. Second, independently of collecting these data,~\cite{cairns1994} also conducted ``direct observation of social interactions among children'' (p. 107), finding that all children in the classroom belonged to a peer group, and that the classroom contained five distinct peer groups. This observational data provides a criterion against which to judge SCM's validity.

In this and subsequent studies, we focus on one statistic of interest, $P$, the proportion of children that SCM identifies as a member of a peer group composed of at least 3 children.\footnote{The requirement that peer groups contain at least 3 members is common in the developmental psychology literature~\citep{shi2014, zhao2016,zarbatany2019}} In this benchmark classroom, based on the independent observational data that all children are members of a peer group, if SCM can detect true positives then $P$ should equal 1.

\subsection{Results}
\begin{center}
\begin{figure}
\includegraphics[width=.5\textwidth]{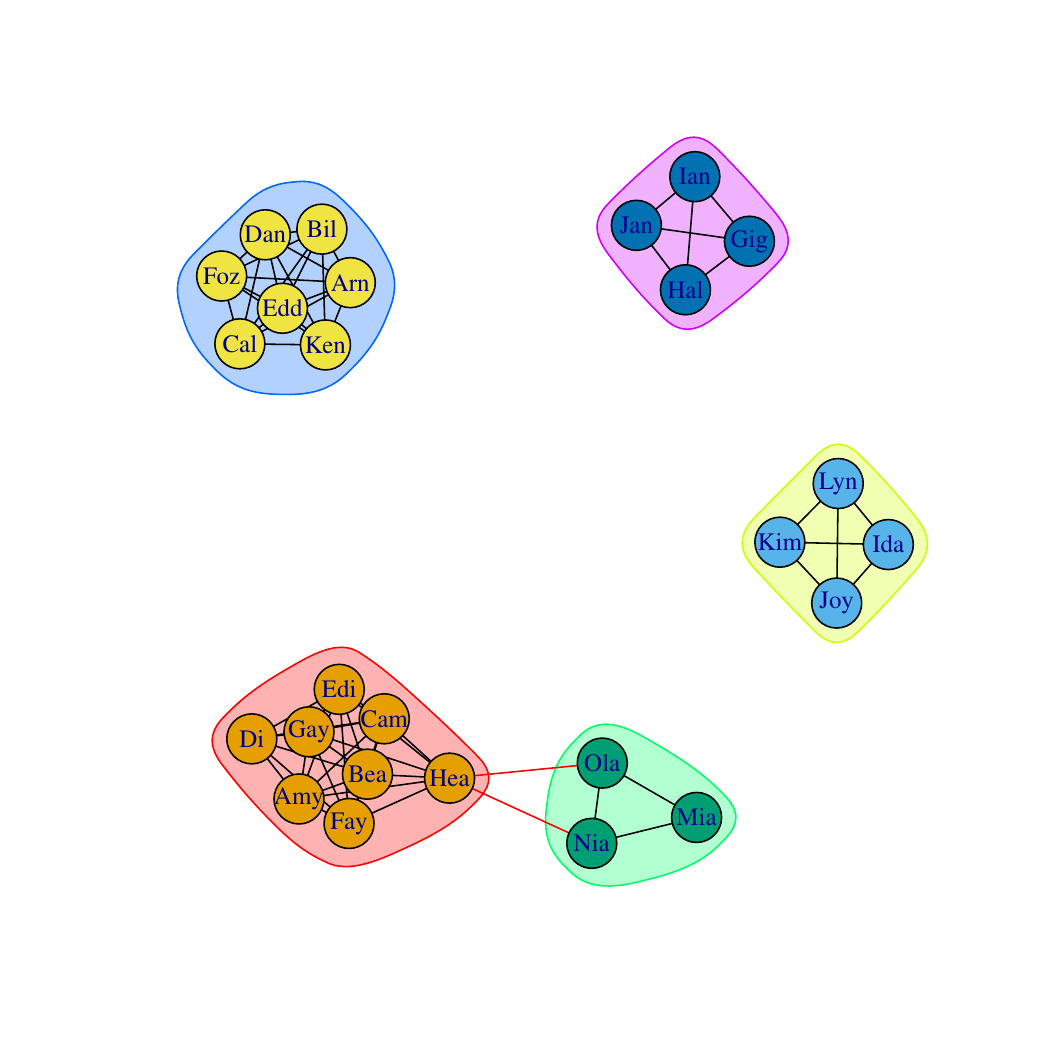}
\caption{Applying SCM to peer report data from a benchmark 7\textsuperscript{th} grade classroom.}
\label{study1}
\centering
\end{figure}
\end{center}

Figure~\ref{study1} shows the peer network obtained by applying SCM to data from a benchmark 7\textsuperscript{th} grade classroom, while the shaded regions outline the peer groups identified by SCM. The identified groups match what ~\cite{cairns1994} observed: cohesive groups of 4 girls, 4 boys, and 7 boys, as well as a larger cluster of 7 girls and 3 girls that are bridged by Heather, who belongs to both groups but here is shown as a member of the larger group. Because all children are identified by SCM as a member of a peer group, $P = 1$, which matches our expectation based on observational data and confirms that SCM detects the true positives in this classroom. From this, we conclude that at least in this benchmark classroom, SCM is able to detect true positives with respect to whether or not children are members of peer groups.

\section{Study 2: False positives in a benchmark classroom}
\subsection{Methods}
Evaluating SCM's risk of false positives is more challenging than examining its ability to detect true positives. In this context, a false positive occurs when SCM identifies children as a members of peer groups when the peer report data lacks evidence of such peer group membership. The most extreme example of peer report data that lacks evidence of peer groups is random peer report data, which a researcher might obtain if the responding children simply guess about peer groups, do not take the data collection seriously and report nonsense, or lack any perceptual accuracy. Although obtaining random peer report data may be unlikely in practice, it provides the most conservative test of SCM's risk of false positives because it is the case where its risk of false positives should be smallest. 

To test SCM's false positive rate, we generated 1000 simulated recall matrices. In each simulated recall matrix, we randomize which children have been reported by their peers to be members of which groups. However, for the sake of comparability and to ensure these simulated data sets are plausible, each simulated recall matrix preserves some features of the original 7\textsuperscript{th} grade classroom described above~\citep{strona2014}. First, the simulated data contain the same number of children (i.e. 26) and the same number of peer reports (i.e. 61). Second, they preserve the salience of each child. For example, in both the original~\cite{cairns1994} data and in every simulated dataset Arn is a high-salience child who was named in 15 peer reports, while Ken is a low-salience child who was named in only 3 peer reports. Finally, they preserve the sizes of the peer reported peer groups. For example, in both the original~\cite{cairns1994} data and in every simulated dataset 28 of the 61 peer reported peer groups contained 4 children, while 1 contained 12 children. This approach yields simulated recall matrices that have many of the same features as the original 7\textsuperscript{th} grade classroom, except that by randomizing which children are reported as members of which groups, should contain no evidence of actual peer groups.

We then use \textsf{SCM 4.0} to identify peer groups from each of these 1000 simulated recall matrices. In each case, we compute our statistic of interest, $P$, the proportion of children that SCM identifies as a member of a peer group composed of at least 3 children. If SCM yields true negatives, correctly failing to identify children as members of peer groups, for which we know these data contain no evidence because they are random, then $P \approx 0$. In contrast, if SCM yields false positives, incorrectly identifying children as members of peer groups, then $P \gg 0$.

\subsection{Results}
\begin{center}
\begin{figure}
\includegraphics[width=.5\textwidth]{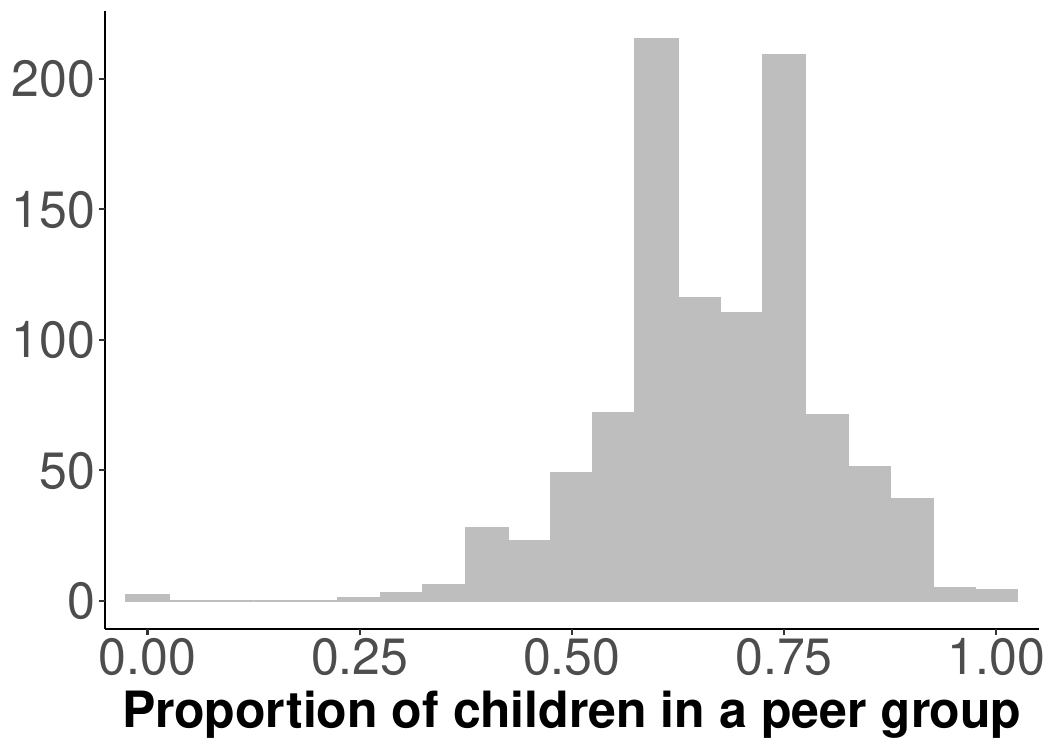}
\caption{False positives applying SCM in a hypothetical 7\textsuperscript{th} grade classroom.}
\label{study2}
\centering
\end{figure}
\end{center}

Figure~\ref{study2} summarizes the value of $P$ obtained by using SCM to identify peer groups in 1000 random recall matrices with characteristics similar to the 7\textsuperscript{th} grade classroom originally studied by~\cite{cairns1994}. These results can be used to answer two questions about SCM. First, how often does SCM yield false positives when applied in a classroom setting like that originally observed by \cite{cairns1994} (i.e. how often is $P > 0$)? We find that $P > 0$ in all 1000 simulated datasets, and therefore that SCM \emph{always} yields false positives in such a setting. Second, how severe are the false positives when applied in this a setting like this? We find that on average SCM assigns two-thirds (M = 0.67, SD = 0.12, min = 0.27, max = 1) of children to a peer group, when in fact the recall matrix contains no evidence of peer groups because it is random.

\section{Study 3: False positives in other classroom}
\subsection{Methods}
In study 2 we investigated SCM's risk of false positives in simulated classrooms that were similar to the benchmark classroom originally studied by~\cite{cairns1994}. However, classrooms can vary widely, and SCM's risk of false positives may be more or less severe in certain types of classrooms. To investigate this possibility, we generated an additional 1000 random recall matrices, varying five features of the simulated classroom: (1) the number of children in the classroom, (2) the number of peer reports provided, (3) the probability that a child is named in a peer report, (4) the amount of skew in the number of times children were named in a peer report, and (5) the amount of skew in the number of children named in a peer report. In this context, skew in the number of times children were named is a measure of child salience; when skew is positive, this corresponds to a classroom where a few children are highly salient and receive many nominations, but most receive few nominations. Skew in the number of children named in a peer report is a measure of group size; when skew is positive, this corresponds to a classroom where some reported groups are large, but most are small. Then, following the same process as study 2, we use \textsf{SCM 4.0} to identify peer groups from each recall matrix and compute our statistic of interest, $P$. Finally, we estimate a regression to examine how characteristics of classroom settings are associated with $P$.

\subsection{Results}
\begin{center}
\begin{figure}
\includegraphics[width=.5\textwidth]{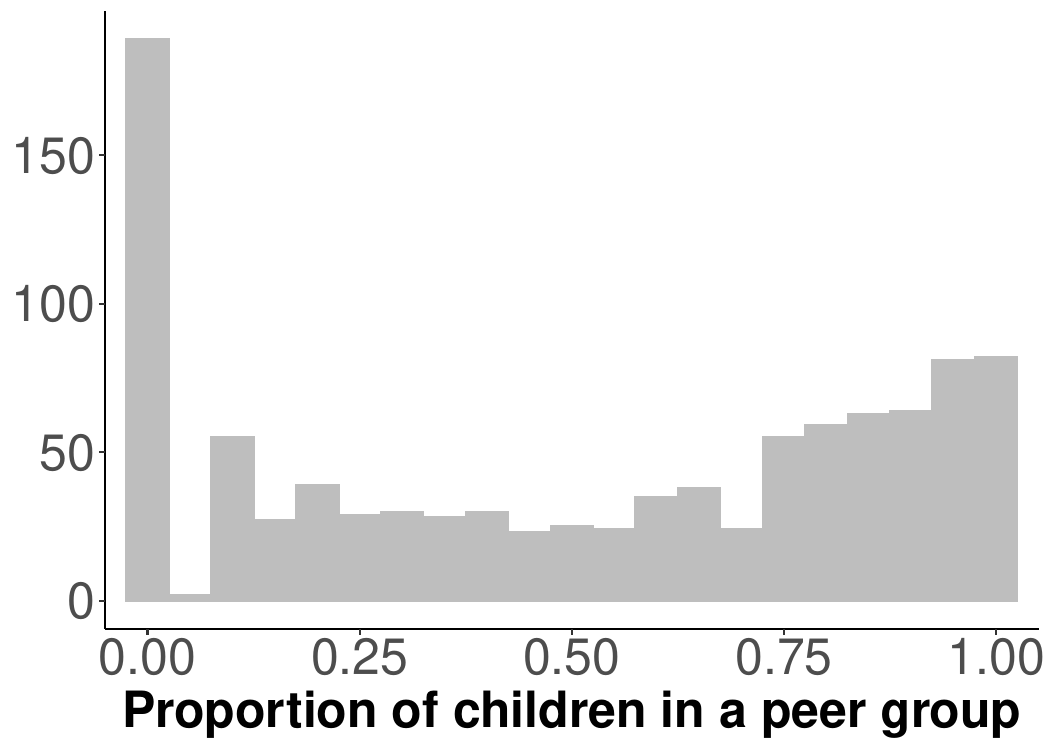}
\caption{False positives applying SCM in classrooms with varying characteristics.}
\label{study3}
\centering
\end{figure}
\end{center}

Figure~\ref{study3} summarizes the value of $P$ obtained by using SCM to identify peer groups in 1000 random recall matrices with the characteristics of classrooms that differ in size, density, child salience, and reported group size. These results can be used to answer two questions about SCM. First, how often does SCM yield false positives when applied in different types of classroom settings (i.e. how often is $P > 0$)? We find that across all settings $P > 0$ in 81.2\% of the 1000 simulated datasets, and therefore that SCM \emph{frequently} yields false positives in a range of classroom settings. Second, how severe are the false positives when applied in this a setting like this? Among datasets where SCM identfied peer groups, it assigned nearly two-thirds (M = 0.63, SD = 0.30, min = 0.075, max = 1) of children to a peer group, when in fact the recall matrix contains no evidence of peer groups because it is random.

Table~\ref{study3_table} reports the results of a regression predicting $P$ as a function of the classroom characteristics that we varied in these datasets, as well as the range of these classroom characteristics across the 1000 simulated datasets. We find that in random data SCM assigns more children to peer groups (i.e. yields more false positives) when the classroom is larger, when children are more likely to be reported as group members, when some children are highly salient, and when some reported groups are large. We also find that SCM assigns fewer children to peer groups when participating children provide more group reports. These estimates are concerning because they highlight that whether SCM assigns a child to a peer group is not based only on patterns in the recall matrix, but is also driven by unrelated characteristics of the data. Moreover, some of the characteristics associated with false positives are precisely the challenges for which SCM was developed to overcome: larger settings where direct network data collection is impractical, and limited numbers of group reports due to low rates of parental and child consent to participate.

\begin{table}
\begin{center}
\begin{tabular}{ p{50mm} c c c | c c } 
 \hline
 Classroom characteristic & b & se & $\beta$ & min & max \\ 
 \hline \hline
 Intercept & 0.252 & 0.043 & --- & --- & --- \\ 
 Number of children & 0.021 & 0.001 & 0.413 & 15 & 40 \\ 
 Number of group reports & -0.008 & 0.000 & -0.880 & 15 & 200 \\ 
 Probability of nomination & 0.013 & 0.001 & 0.189 & 10 & 45 \\ 
 Skew in nominations & 0.100 & 0.015 & 0.115 & -1.77 & 1.99 \\ 
 Skew in reported group size & 0.095 & 0.016 & 0.114 & -0.52 & 2.37 \\ 
 \hline
 \multicolumn{5}{l}{\footnotesize{$R^2 = 0.7147$; P-values are not presented because the datasets are simulated.}}
\end{tabular}
\caption{Impact of classroom characteristics on the number of false positives from SCM.}
\label{study3_table}
\end{center}
\end{table}

\section{Study 4: Backbone extraction and community detection as an alternative}
\subsection{Methods}
Studies 2 and 3 suggest that SCM has a very high risk of false positives, identifying children as members of peer groups even in random data, under a wide range of circumstances. In this study, we explore one potential alternative to identifying peer groups from peer report data: backbone extraction and community detection.

Backbone extraction methods offer an alternative way to transform a recall matrix $\mathbf{R}$ into a binary peer network $\mathbf{N}$~\citep{neal2014}. These methods begin similar to SCM by transforming data such as a recall matrix (known as a \emph{bipartite} matrix) into a co-occurrence matrix (known as a bipartite \emph{projection}). Unlike SCM, backbone extraction then applies a statistical test to determine which pairs of children co-occur enough times to infer that they likely socially interact, and constructs a peer network using these statistically significant links. There are several methods for performing backbone extraction, but here we focus on the Stochastic Degree Sequence Model (SDSM) because it is fast, well-documented, and easy to compute using the \textsf{R} \texttt{backbone} package~\citep{backbone}.

Community detection methods offer an alternative way to identify peer groups from a binary peer network. These methods aim to identify cohesive groups in a network such that the majority of relationships are located within group and few relationships are located between groups. There are several methods for identifying these groups, including methods that allow children to have multiple group memberships, and methods that allow groups to have fuzzy boundaries~\citep{fortunato2010}. In this study we use a method known as modularity maximization because it is among the most commonly used in network analysis, it identifies the \emph{optimal} way to assign children to groups, and it is easy to compute using the \textsf{R} \texttt{igraph} package~\citep{igraph}. Like SCM, this method identifies groups with distinct rather than fuzzy boundaries, but unlike SCM it requires group memberships to be mutually exclusive. While this is an important difference, in practice it may play a limited role because most studies using SCM already focus only on each child's one primary peer group \citep{berger2012,chung2010, zarbatany2019, zhao2016}. 

In this study, we repeat studies 1 - 3 using a combination of backbone extraction and community detection (BE-CD) to identify peer groups rather than SCM. Before turning to the results, we briefly illustrate how BE-CD can be used in the \textsf{R} software. The first time BE-CD is used, the \texttt{backbone} and \texttt{igraph} packages must be installed in \textsf{R} by typing:
\begin{verbatim}
    install.packages("backbone")
    install.packages("igraph")
\end{verbatim}
These two packages must be loaded so that \textsf{R} can use them by typing:
\begin{verbatim}
    library(backbone)
    library(igraph)
\end{verbatim}
If the recall matrix is stored as a CSV file called \texttt{recall.csv}, then the BE-CD approach to identifying peer groups involves typing:
\begin{verbatim}
    R <- read.csv(recall.csv)
    N <- sdsm(R)
    N <- backbone.extract(N, signed=F)
    N <- graph_from_adjacency_matrix(N, diag=F,
         mode="undirected")
    N.groups <- cluster_optimal(N)
\end{verbatim}
This series of five commands imports the recall matrix data, conducts the SDSM statistical test, extracts the peer network, converts the network into a form that \texttt{igraph} can understand, and identifies peer groups. After these commands, the results can be examined using:
\begin{verbatim}
    membership(N.groups)
    plot(N.groups, N)
\end{verbatim}
The first command will show the peer group membership of each child, while the second command will plot the peer network and show the boundaries of the peer groups.

\subsection{Results}
\begin{center}
\begin{figure}
\includegraphics[width=\textwidth]{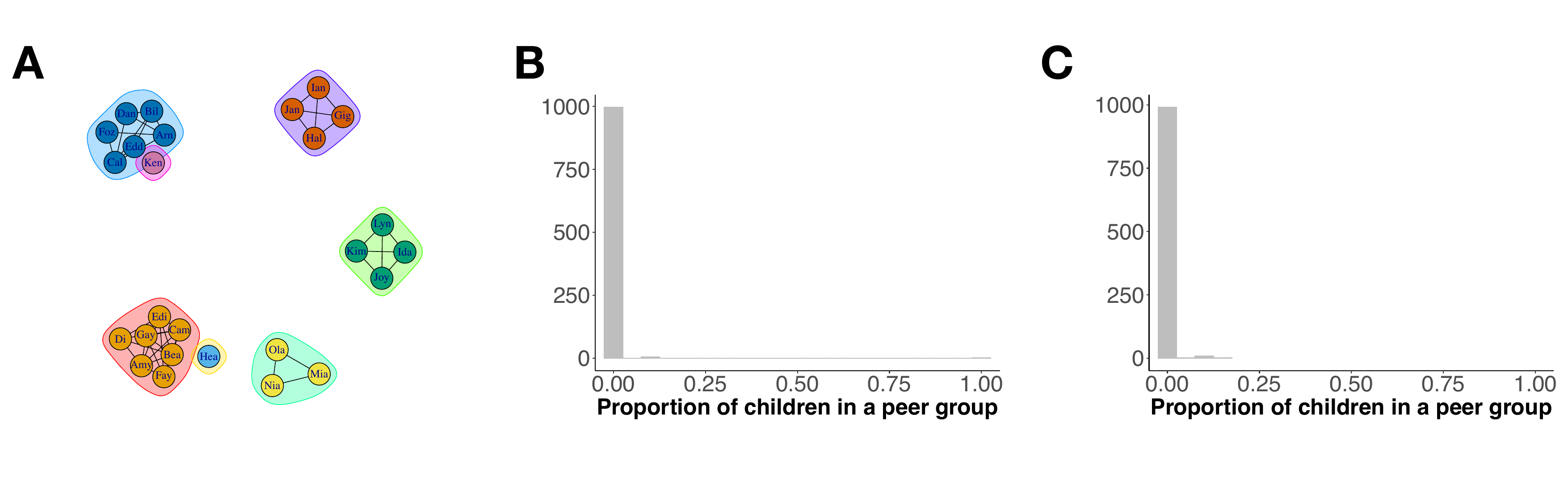}
\caption{Identifying peer groups using backbone extraction and community detection (BE-CD). (\textbf{A}) Applied to \cite{cairns1994} 7\textsuperscript{th} grade classroom; (\textbf{B}) Applied to 1000 simulated 7\textsuperscript{th} grade classrooms; (\textbf{C}) Applied to 1000 classrooms with varying characteristics.}
\label{study4}
\centering
\end{figure}
\end{center}

Figure~\ref{study4}A replicates study 1 by using BE-CD to identify peer groups in the benchmark data described by~\cite{cairns1994} (c.f. Figure~\ref{study1}). The identified peer groups almost perfectly match those identified by SCM, and those described by~\cite{cairns1994} from their direct observations. However, there are two exceptions: Heather and Ken are not assigned to peer groups. These exceptions offer an opportunity to compare the tuning accuracy of SCM and BE-CD by considering whether the recall matrix contains sufficient evidence to believe Heather and Ken are members of peer groups (consistent with SCM), or insufficient evidence than Heather and Ken are members of peer groups (consistent with BE-CD). Peer reports about Heather's hanging out behaviors were mixed: Four children reported that Heather was a member of the larger group of girls only, another four reported she was a member of the smaller group only, and no one reported she was a member of both groups. SCM views this as sufficient evidence to conclude that Heather is a member of both groups, while BE-CD views this evidence as too mixed to conclude she is a member of either group. Ken offers a similarly ambiguous case: only three children reported about Ken's hanging around behavior: two reported he hangs around with the larger group of boys, while one reported he hangs around with the smaller group of boys. SCM views this as sufficient evidence to definitively conclude Ken is linked to all the boys in the larger group, and none in the smaller group, while BE-CD again finds the evidence too mixed to draw a conclusion. We are unable to determine whether SCM or BE-CD is `right,' but based on~\cite{cairns1994} description of their observations and on the groups reported by the children, both seem plausible.

Figure~\ref{study4}B replicates study 2 by using BE-CD to identify peer groups in 1000 randomized versions of~\cite{cairns1994} data (c.f. Figure~\ref{study2}). These results can be used to ask: how often does BE-CD yield false positives when applied in a classroom setting like that originally observed by \cite{cairns1994} (i.e. how often is $P > 0$)? We find that BE-CD yields false positives in only 5 of the 1000 datasets. Among those few cases where it did yield false positives, they were not severe, assigning only 11.5\% of children to a peer group when the correct value is 0\%.

Figure~\ref{study4}C replicates study 3 by using BE-CD to identify peer groups in 1000 classrooms with varying characteristics (c.f. Figure~\ref{study3}). Again, these results can be used to ask: how often does BE-CD yield false positives when applied in a classroom setting like that originally observed by \cite{cairns1994} (i.e. how often is $P > 0$)? We find that BE-CD yields false positives in only 10 of the 1000 datasets. Among those cases where it did yield false positives, they were not severe, assigning between 7.5\% and 13.6\% of children to a peer group when the correct value is 0\%.

\section{Discussion}
Social cognitive mapping (SCM) is a method for identifying peer groups from peer report data. As formalized by~\cite{cairns1994} and implemented in \textsf{SCM 4.0}~\citep{leung1998}, it is the most common method for identifying peer groups in developmental psychology, and is widely used in other fields including school psychology, social psychology, special education, and substance use. In study 1, we found that SCM can identify peer groups that are known from direct observation to actually exist (true positives). However, in studies 2 and 3, we found that SCM also frequently identifies peer groups that are known to \emph{not} exist (false positives). The results from study 3 also demonstrates that the severity of false positives is greatest in those settings where SCM was designed to be used: larger classrooms with lower participation rates. Finally, in study 4, we introduced backbone extraction and community detection as an alternative to SCM, and found that it has a similar ability as SCM to detect true positives, but a much lower risk than SCM of detecting false positives.

Based on our detailed review of SCM, the associated \textsf{SCM 4.0} program, and the results of these four studies, we offer four recommendations to developmental psychologists and others wishing to identity peer groups from peer report data. First, because SCM as implemented in the \textsf{SCM 4.0} program has a very high risk of false positives and because key parts of the \textsf{SCM 4.0} are undocumented, \emph{researchers should not use the \textsf{SCM 4.0} program}. Second, and for the same reasons, findings about peer groups reported in \emph{papers using \textsf{SCM 4.0} should be viewed with caution}. Third, because multiple variants of SCM exist~\cite[e.g.,][]{cairns1994,kindermann1993,gest2007}, \emph{researchers using SCM should be explicit and detailed about the exact procedures they employ}, including reporting the cut-off threshold and the method for identifying peer groups from a binary network. Finally, although in 1994 ``the scientific assessment of peer groups [was limited] by a gap in methods available for social network analysis''~\citep[][p. 100]{cairns1994}, this is no longer the case. Therefore, when seeking to identify peer groups from peer report data, \emph{researchers should consider using well-documented and statistically-informed network analytic methods}. In this paper we have illustrated how backbone extraction and community detection (BE-CD) might be used. However, many alternatives to SCM are now available.

In addition to offering researchers concrete recommendations for identifying peer groups from peer report data, this work also highlights several opportunities for future research. First, we have focused on evaluating the risk of false positives in the most widely-used variant of SCM~\citep{cairns1994}, but future studies should investigate the risk of false positives in other variants of SCM~\citep[e.g.][]{kindermann1993,gest2007}. Second, we have focused only on tuning accuracy, however future studies should investigate the perceptual accuracy and validity of peer group identification methods. Such studies will be challenging because they will require identifying peer groups through direct behavioral observations, which are challenging to collect~\citep{gest2003}. Investigation of perceptual accuracy would evaluate whether the peer groups reported to exist by participating children mirror observed peer groups, while investigation of validity would evaluate whether the peer groups identified by pooling such peer report data mirror observed peer groups.

Studies of peer groups identified via SCM, and in particular \textsf{SCM 4.0}, remain common in the developmental literature. However, SCM has an unacceptably high rate of false positives, casting doubt on whether the peer groups it identifies actually exist. Because understanding peer groups remains essential for understanding a wide range of developmental processes, developmental researchers must adopt alternative methods for identifying peer groups.

\textbf{Conflict of Interest Statement:} The authors have nothing to disclose.

\end{document}